# Monitoring of Low Voltage Distribution Grid Considering the Neutral Conductor

Andreas Kotsonias, *Student Member, IEEE,* Lenos Hadjidemetriou, *Member, IEEE,* Markos Asprou *Member, IEEE,* and Elias Kyriakides, *Senior Member, IEEE*

***Abstract*—The most widely used method for monitoring Low Voltage Distribution Grids (LVDGs) is the three phase Weighted Least Squares (WLS) State Estimation (SE), which was initially developed for Medium Voltage Distribution Grids (MVDGs). This methodology is implicitly applied in LVDGs with an assumption that the neutral conductor represents a zero potential across the whole system. However, this assumption is often not valid for LVDGs as the consumers' loads are highly asymmetrical and the neutral conductor is usually grounded only at the MV-LV transformer substation. Therefore, if this method is applied for the monitoring of LVDGs it may deteriorate the performance of the WLS SE leading to inaccurate results. In this paper, an investigation is initially conducted in order to evaluate the performance of the conventional WLS SE methodology when applied for the monitoring of LVDGs. The results of this study indicate that the performance of this methodology is not consistent and is highly affected by the operating conditions of the LVDG. In view of these results, a new methodology is proposed and developed that takes into consideration the effect of the neutral conductor and as a result an outstanding performance is achieved compared to the conventional methodology.**

## I. Introduction

THE need for highly decarbonized power systems requires significant restructuring of their infrastructure and adjustments in their operation. In this attempt, the LVDGs are already undergoing vast structural and operational changes compared to the high and medium voltage levels of the system. Traditionally, LVDGs were characterized as highly passive systems with an understandable interaction between the grid and the consumers. The main factors behind the rapid development of LVDGs is the massive integration of Distributed Energy Resources (DERs), in the form of rooftop residential photovoltaic systems (PV) and energy storage devices [1], the electrification of cooling, heating and transportation sector (electric vehicles), as well as the envisioned demand response schemes [2]. These aforementioned factors are transforming LVDGs from passive systems into highly complex and active systems with fast dynamic behavior. Therefore, there is a need for exploiting all the available flexibilities in novel control schemes for LVDGs [3] in order to support their revolution and to maintain a secure and reliable operation with a high penetration level of DERs. An important aspect for the effective operation of any control scheme is an accurate and reliable monitoring system.

The monitoring of power systems has been an important topic for decades, with the researchers and industry mainly focusing on transmission systems. In [4]-[5] the widely used and popular WLS SE is presented. Due to its proven good results and relatively low complexity, various alternatives of WLS SE have been proposed in an effort to further increase its performance with their main difference being the chosen state variables [6]. MVDGs have also been a focus area of the research efforts. Due to different characteristics compared to transmission systems (i.e., unbalanced operation, distribution lines with higher *R*/*X* ratios, and radial topology instead of meshed), the monitoring methods developed for transmission systems cannot be applied directly to MVDGs. In order to account for these characteristics, a three phase coupled WLS SE is proposed in [7] which is extended in [8] to also cover the modeling of the various system components. An important aspect of this method is that it uses Carson's equations [9] in order to construct the impedance matrix for each distribution line. Considering that the most common transformer configuration in MV-LV substations is D-$Y_g$, the zero sequence component of the asymmetrical currents inside the LVDG is eliminated and does not propagate in the MVDG. Therefore, the current flow in the neutral conductor of MVDGs (if such a conductor exists) is negligible and the voltage drop across it can be approximated to zero. This allows using Kron's reduction [7]-[9], which reduces the impedance matrix to 3x3, simplifying the monitoring of MVDGs.

On the contrary, the research effort regarding appropriate monitoring methods in LVDGs was very limited (up to recently). The main factors behind this gap is not only due to their traditional passive behavior but also due to insufficient measuring equipment and infrastructure at the low voltage level. The latter makes LVDGs unobservable [5] and consequently limits the capabilities of appropriate monitoring schemes. However, with the recent large scale deployment of smart meters [10]-[11], new opportunities arise. The main reason of their commissioning is for billing purposes and in general for understanding customers' behavior, but their use can simultaneously enable the monitoring of LVDGs, if they are utilized properly. For that purpose, the monitoring of LVDGs started gaining research attention by utilizing the

This research was supported by the Electricity Authority of Cyprus (EAC). EAC is a partner of the Innovation Hub of the KIOS Center of Excellence, which is enabled by the EU's strategic Horizon 2020 program for "Spreading Excellence and Widening Participation – Teaming."

A. Kotsonias, L. Hadjidemetriou, M. Asprou and E. Kyriakides are with the KIOS Research and Innovation Center of Excellence and the Department of Electrical and Computer Engineering, University of Cyprus, 1678 Nicosia, Cyprus (e-mail: kotsonias.andreas@ucy.ac.cy, hadjidemetriou.lenos@ucy.ac.cy asprou.markos@ucy.ac.cy, ac.cy, elias@ucy.ac.cy).

capabilities of the smart meters. In [12], their technological advancements are presented as well as a monitoring system for an LVDG with high PV penetration that utilizes 48 smart meters. Similarly, in [13] the WLS SE methodology is enabled by the use of smart meters for the monitoring of an LVDG with nodal voltages expressed in polar form as the chosen state variables. In [14], a new scheme for assigning measurement weights is proposed that takes into consideration the time skewness of the smart meter measurements. This work is extended in [15] by taking into account the possible consumers' load variation between updating intervals of the smart meters. In [16], an analysis is conducted regarding the performance of the three phase WLS SE when nodal voltages or branch currents are used as the state variables. In both case studies the state variables were expressed in their polar and rectangular forms and from the results it can be concluded that regardless the plane that the state variables are expressed into, both estimators yield identical results. A cloud based architecture is proposed in [17] that divides an LVDG into blocks, with each block being observable by smart meters that report to a concentrator. Despite the drawback of requiring a concentrator for each block, this setup allows the parallel execution of the WLS SE in each block by utilizing the computing capabilities of the concentrators. As a result, the overall estimation of the system state requires significantly lower computational time. In [18], a monitoring solution is presented that avoids overloading the control center with unnecessary measurement information by transferring some of the data analysis and decision making procedures at the secondary substation. Additionally, this solution offers real time monitoring of an LVDG by using smart meters and substation measurements to enable a branch current based WLS SE. A multi-area SE is proposed in [19] as well as a clear definition of the necessary hardware and software requirements to achieve real time monitoring based on substation automation units.

A common attribute of the aforementioned works, is that the conventional three phase WLS SE (C-WLS SE), which was developed for MVDGs, was applied for monitoring LVDGs and as a result it was implicitly assumed that MVDGs and LVDGs have similar characteristics. However, the assumption required by the C-WLS SE that the neutral conductor's voltage is at zero potential is not always valid in LVDGs. In the contrary, most suburban and rural LVDGs (especially in Europe) have their neutral conductor grounded only at the substation transformer. Hence, unlike MVDGs, the asymmetrical loading conditions of LVDGs are causing a significant current flow through the neutral conductor (zero sequence current), which creates a voltage drop across it. Therefore, by making the assumption that the neutral conductor is at zero potential across the whole system and using Kron's reduction to simplify the system model can lead to inaccurate results in the monitoring of LVDGs. More specifically, this assumption can affect the accuracy of C-WLS SE in two distinct ways:

1) The system model that will be used for the SE procedure will not be an accurate representation of the actual system and hence the estimated state may have significant error.
2) The voltage measurements that the smart meters provide are phase to neutral measurements [20]. Since the voltage of the neutral conductor can have different values at different locations, the voltage measurements that are used in the SE procedure are expressed with a different reference voltage.

Thus, the aim and contribution of this paper is firstly to investigate how the accuracy of C-WLS SE is affected by the specific characteristics and operating conditions of LVDGs. For this purpose, numerous scenarios are conducted by applying the C-WLS SE in the monitoring of the IEEE European Low Voltage Test Feeder. Moreover, it is assumed that the system is observable by smart meters and that the neutral conductor is grounded only at the Y side of the D-$Y_g$ transformer at the MV-LV substation. Secondly, a new and innovative WLS SE method is proposed, referred to as N-WLS SE, that takes into account the full effect of the neutral conductor by including the neutral voltage in the state vector. In a benchmarking investigation between the proposed N-WLS SE and the conventional C-WLS SE, the proposed method presents an outstanding performance regarding the monitoring accuracy under any characteristics and any operating conditions of the LVDG. The paper is organized as follows. In Section II, the theory of the C-WLS SE is presented while in Section III the investigation of its performance is conducted by applying it in the monitoring of the IEEE European Low Voltage Test Feeder. In Section IV, the new methodology is introduced and its performance is compared with the C-WLS SE. Finally, the paper concludes in Section V.

## II. Conventional Three Phase WLS SE

State estimation is a well known technique in power systems which has been successfully applied for decades and it's an essential part of SCADA and its functionalities (i.e., voltage control, economic dispatch, optimal power flow, etc.). By maximizing the likelihood probability of having a given set of measurements, WLS SE produces the best estimate for the system's state that minimizes the weighted measurement errors [5]. A given set of measurements can be represented as,

$$z = h(x) + e \tag{1}$$

where $z$ is a vector containing the available measurements, $h(x)$ is a vector containing nonlinear functions that relate mathematically the state variables with the measurements, $x$ is the state vector and $e$ is the measurement error vector. In WLS SE it is assumed that the measurement errors follow a normal distribution with zero mean and known variance [6]-[7], i.e. $e \sim N(0, \sigma^2)$.

### *A. The Weighted Least Squares State Estimation Procedure*

In this section the main procedure and equations of the three phase SE are outlined in order to set up the problem and introduce the modifications required by the proposed method.

According to the WLS SE framework, the state vector can be derived by minimizing the function $J(x)$ of (2), which corresponds to the summation of the squared measurement residuals that are weighted by the error variance of the measurement device [5],

$$\min_x J(x) = \sum_{i=1}^{n} \frac{[z_i - h_i(x)]^2}{\sigma_i^2} = [z - h(x)]^T R^{-1} [z - h(x)] \tag{2}$$
$$Subject\ to\ c(x) = 0$$

where $n$ is the number of measurements and $R$ is the measurement error covariance matrix. Under the framework of C-WLS SE, it is assumed that there is no correlation between the metering errors [21] and as a result $R$ is a diagonal matrix ($R = diag\{\sigma_1^2,\ \sigma_2^2, \dots, \sigma_n^2\}$, where $\sigma_i$ is the standard deviation associated with the $i$-th measurement). Zero injection nodes have virtual PQ measurements with very high measurement weights and as a result, in order to avoid ill conditioning, the problem is formulated as an optimization problem with equality constraints [22]. These constraints represent the zero injection measurements and are related to the state variables by nonlinear functions in vector $c(x)$. Optimization problems with equality constraints are usually solved by applying the Langrangian method. In this occasion however, due to the use of nonlinear functions in vectors $h(x)$ and $c(x)$, it is necessary to also apply the Gauss-Newton iterative process. Hence, the C-WLS SE requires to solve the below iterative scheme,

$$\begin{bmatrix} \Delta x \\ \lambda \end{bmatrix} = \begin{bmatrix} H^T(x^k)R^{-1}H(x^k) & -C^T(x^k) \\ C(x^k) & 0 \end{bmatrix}^{-1} \begin{bmatrix} H(x^k)R^{-1}\Delta z^k \\ -c(x^k) \end{bmatrix} \quad (3)$$

where $k$ is the iteration number, $\Delta x = x^k - x^{k+1}$, $\Delta z = z - h(x^k)$, $\lambda$ is the Lagrangian multipliers vector, $H(x^k) = \left.\frac{\partial h(x)}{\partial x}\right|_{x^k}$, $C(x^k) = \left.\frac{\partial c(x)}{\partial x}\right|_{x^k}$

### *B. Measurement Functions*

In LVDGs, the available measurements that can be used in the SE procedure are mainly the active and reactive power consumption of the various consumers, as well as the voltage magnitude at their premises [15]. Therefore, the measurement vector $z$ of C-WLS SE is formed as,

$$z = \left[P_{inj}\ Q_{inj}\ |V|\right]^T \quad (4)$$

These measurements are related mathematically to the state variables (for this paper without losing generality will be the nodal voltages in polar form) with nonlinear functions in vectors $h(x)$ and $c(x)$. Moreover, due to high asymmetry and coupling between phases, the WLS SE cannot be conducted using a single-phase equivalent approach as in transmission systems [7]. For distribution grids, WLS SE must consider all phases simultaneously; therefore each three phase node is characterized by six state variables. The nonlinear functions that are included in vectors $h(x)$ and $c(x)$ are more complex in three phase analysis as they also contain cross product terms between phases for the active and reactive power injections [6],

$$P_i^p = |V_i^p| \sum_{k=1}^{n} \sum_{m=a}^{c} |V_k^m| \left(G_{ik}^{pm} cos\delta_{ik}^{pm} + B_{ik}^{pm} sin\delta_{ik}^{pm}\right) \quad (5)$$

$$Q_i^p = |V_i^p| \sum_{k=1}^{n} \sum_{m=a}^{c} |V_k^m| \left(G_{ik}^{pm} sin\delta_{ik}^{pm} - B_{ik}^{pm} cos\delta_{ik}^{pm}\right) \quad (6)$$

where $n$ is the total number of nodes, $P_i^p$ and $Q_i^p$ is the injected active and reactive power respectively at node $i$ in phase $p$, $\delta_{ik}^{pm} = \delta_i^p - \delta_j^m$ and $Y_{ik} = G_{ik} + jB_{ik}$ is the element of the admittance matrix of the system. For a system with two nodes $i$ and $k$,

$$Y = \begin{bmatrix} Y_{ii} & Y_{ik} \\ Y_{ki} & Y_{kk} \end{bmatrix} \quad (7)$$

where each element of the $Y$ matrix represents a 3x3 submatrix, i.e., $Y_{ii}$ is equal to,

$$Y_{ii} = \begin{bmatrix} Y_{ii}^{aa} & Y_{ii}^{ab} & Y_{ii}^{ac} \\ Y_{ii}^{ba} & Y_{ii}^{bb} & Y_{ii}^{bc} \\ Y_{ii}^{ca} & Y_{ii}^{cb} & Y_{ii}^{cc} \end{bmatrix} \quad (8)$$

where $Y_{ii}^{aa}$ is the self admittance of phase $a$ at node $i$, $Y_{ii}^{ab}$ is the mutual admittance between phases $a$ and $b$ at node $i$ and etc. (similarly for $Y_{ik}$, $Y_{ki}$ and $Y_{kk}$). The voltage measurements are directly related to their corresponding state variable with $x_i = z_i$, where $x_i$ is the state variable corresponding to a voltage magnitude at a specific node and phase while $z_i$ is the voltage measurement at the same node and phase.

### *C. Feeder Model*

In a low voltage distribution feeder, the main components are usually the distribution lines, the MV-LV transformer and the consumers' loads. Since the purpose of this case study is to investigate the performance of C-WLS SE when applied in the monitoring of LVDGs under steady state conditions, the consumers' loads are modeled as constant impedance elements.

1) *Low Voltage Distribution Lines*: As it is shown in Fig. 1, the distribution lines in LVDGs usually consist of four conductors (three phase conductors and one neutral conductor), with couplings between them. Neglecting the line charging effects, as they are insignificant in the low voltage level, the impedance matrix characterizing this line is [23],

$$Z_{4x4} = \begin{bmatrix} z_{aa} & z_{ab} & z_{ac} & z_{an} \\ z_{ba} & z_{bb} & z_{bc} & z_{bn} \\ z_{ca} & z_{cb} & z_{cc} & z_{cn} \\ z_{na} & z_{nb} & z_{nc} & z_{nn} \end{bmatrix} \quad (9)$$

The elements of the 4x4 impedance matrix are calculated using the modified Carson's equations [9],

$$z_{ii} = r_i + 0.0493 + j0.0628\left(ln\frac{0.3048}{GMR_i} + 8.0251\right)\Omega/km \quad (10)$$

$$z_{ij} = 0.0493 + j0.0628\left(ln\frac{0.3048}{D_{ij}} + 8.0251\right)\Omega/km \quad (11)$$

where,

$z_{ii}$ is the self-impedance of conductor $i$ (Ω/km)
$z_{ij}$ is the mutual impedance between conductor $i$ and $j$ (Ω/km)
$r_i$ is the ac resistance of conductor $i$ (Ω)
$GMR_i$ is the Geometric Mean Radius of conductor $i$ (m)
$D_{ij}$ is the distance between conductor $i$ and $j$ (m)

The expressions in (10) and (11) assume a 50 Hz system and a constant earth resistivity of 100 Ω.m. A more general expression can be found in [23]. The modified Carson's equations are chosen over their full expressions [9] as it has been proven that despite the simplification of the expressions, the end result has an error of less than 0.3% in the construction of the impedance matrix. In MVDGs, the neutral conductor, if it exists, can be grounded at multiple points. Moreover, due to the common D-$Y_g$ transformer connection in the MV-LV substation, there will be no significant current through the neutral conductor. Therefore the voltage drop can be approximated to zero. For this reason, in C-WLS SE it is assumed that the neutral conductor is grounded at both ends of a distribution line (Fig. 1) and that it represents a zero potential across the neutral conductor of the whole system [7], [23]. This assumption allows the use of Kron's reduction (12) in order to eliminate the neutral wire and reduce the 4x4 impedance matrix in (9) to a simpler form of 3x3 (13),

$$z'_{ij} = z_{ij} - \frac{z_{in} \times z_{nj}}{z_{nn}} \tag{12}$$

$$Z_{3x3} = \begin{bmatrix} z'_{aa} & z'_{ab} & z'_{ac} \\ z'_{ba} & z'_{bb} & z'_{bc} \\ z'_{ca} & z'_{cb} & z'_{cc} \end{bmatrix} \tag{13}$$

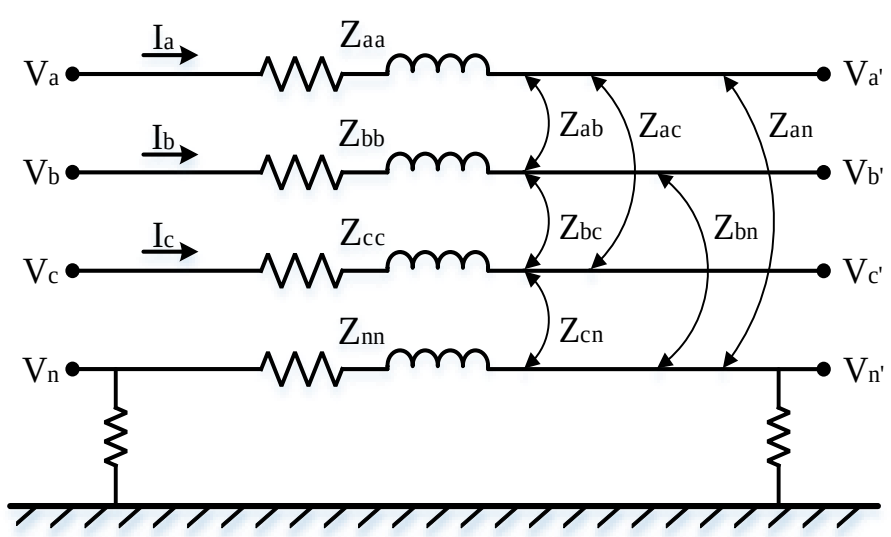

Fig. 1. A four wire neutral grounded distribution line

It should be noted that the use of Kron's reduction in order to represent a 4 wire distribution line with a 3x3 impedance matrix, decreases slightly the self impedances and decreases significantly the mutual impedances.

2) *MV-LV Transformer:* An important element of an LVDG is the step down transformer at the substation. This transformer can have any winding connection, but the most commonly used is D-$Y_g$. Since in the C-WLS SE only the three power phases are considered, the transformer is modeled as an element with six terminals, which is characterized by a 6x6 admittance matrix. The processes of determining this admittance matrix for any winding connection can be found in [24]. Here, only the admittance matrix of a D-$Y_g$ transformer is presented,

$$Y_T = \begin{bmatrix} Y_{pp} & Y_{ps} \\ Y_{sp} & Y_{ss} \end{bmatrix} \tag{14}$$

$$Y_{pp} = \frac{1}{3}\begin{bmatrix} 2 & -1 & -1 \\ -1 & 2 & -1 \\ -1 & -1 & 2 \end{bmatrix} y_t \tag{15}$$

$$Y_{ps} = \frac{1}{\sqrt{3}}\begin{bmatrix} -1 & -1 & 0 \\ 0 & -1 & 1 \\ 1 & 0 & -1 \end{bmatrix} y_t \tag{16}$$

$$Y_{sp} = Y_{ps}^T \tag{17}$$

$$Y_{ss} = \begin{bmatrix} 1 & 0 & 0 \\ 0 & 1 & 0 \\ 0 & 0 & 1 \end{bmatrix} y_t \tag{18}$$

where $Y_T$ is the 6x6 admittance matrix of a D-$Y_g$ transformer and $y_t$ is the per unit transformer leakage admittance.

## III. IEEE European LV Test Feeder Case Study

In this section, the accuracy of the C-WLS SE is investigated when it is applied for monitoring an LVDG with a neutral conductor grounded only at the MV-LV substation. For this investigation, the topology of the IEEE European Low Voltage test feeder is used as illustrated in Fig. 2. It should be noted that this system has been slightly modified to include three phase loads in order to account for all possible load connections. Additionally, since the exact knowledge of the distribution line geometry and conductor parameters are highly significant in order to evaluate the effect of the neutral conductor correctly, it is considered that the main feeder lines are 4x100 $mm^2$ OH aluminum lines, while the supply lines from the main feeder poles to the consumers' premises are considered to be 2x22 $mm^2$ OH aluminum lines. The impedance matrices for both types of distribution lines are provided in Tables I and II.

TABLE I
Impedance Matrix Of 4x100 $mm^2$ OH Aluminum Line

| R (Ω/km) | | | | L (H/km) | | | |
|---|---|---|---|---|---|---|---|
| 0.3187 | 0.0482 | 0.0482 | 0.0482 | 0.0024 | 0.0016 | 0.0014 | 0.0013 |
| 0.0482 | 0.3187 | 0.0482 | 0.0483 | 0.0016 | 0.0024 | 0.0016 | 0.0014 |
| 0.0482 | 0.0482 | 0.3188 | 0.0483 | 0.0014 | 0.0016 | 0.0024 | 0.0016 |
| 0.0482 | 0.0483 | 0.0483 | 0.3188 | 0.0013 | 0.0014 | 0.0016 | 0.0024 |

TABLE II
Impedance Matrix Of 2x22 $mm^2$ OH Aluminum Line

| R (Ω/km) | | L (H/km) | |
|---|---|---|---|
| 1.2753 | 0.0482 | 0.0025 | 0.0016 |
| 0.0482 | 1.2753 | 0.0016 | 0.0025 |

In order to enable the use of the C-WLS SE the system must be observable. Therefore, it is assumed that all consumers are equipped with a smart meter that can provide active and reactive power injection measurements, as well as voltage magnitude measurements for each phase. Additionally, without losing generality, it is assumed that the smart meters belong to the accuracy class 0.5s for active power measurements and to class 1 for reactive power measurements. Furthermore, it is assumed that the measurement errors have a 95% confidence interval [25], i.e. there is a 95% probability that the measurement errors will be within the interval bounded by the maximum errors defined in Table III (based on the accuracy class of the smart meters).

TABLE III
Maximum Measurement Errors

| Type of measurement | Full scale | ±Percentage error limits |
|---|---|---|
| $P_{inj}$ | 9.2 kVA | 0.6% |
| $Q_{inj}$ | 9.2 kVA | 1% |
| \|V\| | 300 V | 0.4% |

The system is implemented in MATLAB/Simulink where the power flow results of each scenario represent ideal measurements and are in fact the true state of the system. The actual measurements that are used in the SE procedure are defined as,

$$z_{meas} = z_{true} + FS \cdot N\left(0, \sigma_{P,Q,|V|}\right) \tag{19}$$

where $z_{true}$ is the true value of the measurement, *FS* is the full scale meter reading associated with each type of measurement and $\sigma_{P,Q,|V|}$ are the standard uncertainties for each measurement type. These standard uncertainties are calculated by dividing the maximum measurement errors defined in Table III with 1.96 (due to the 95% confidence interval).

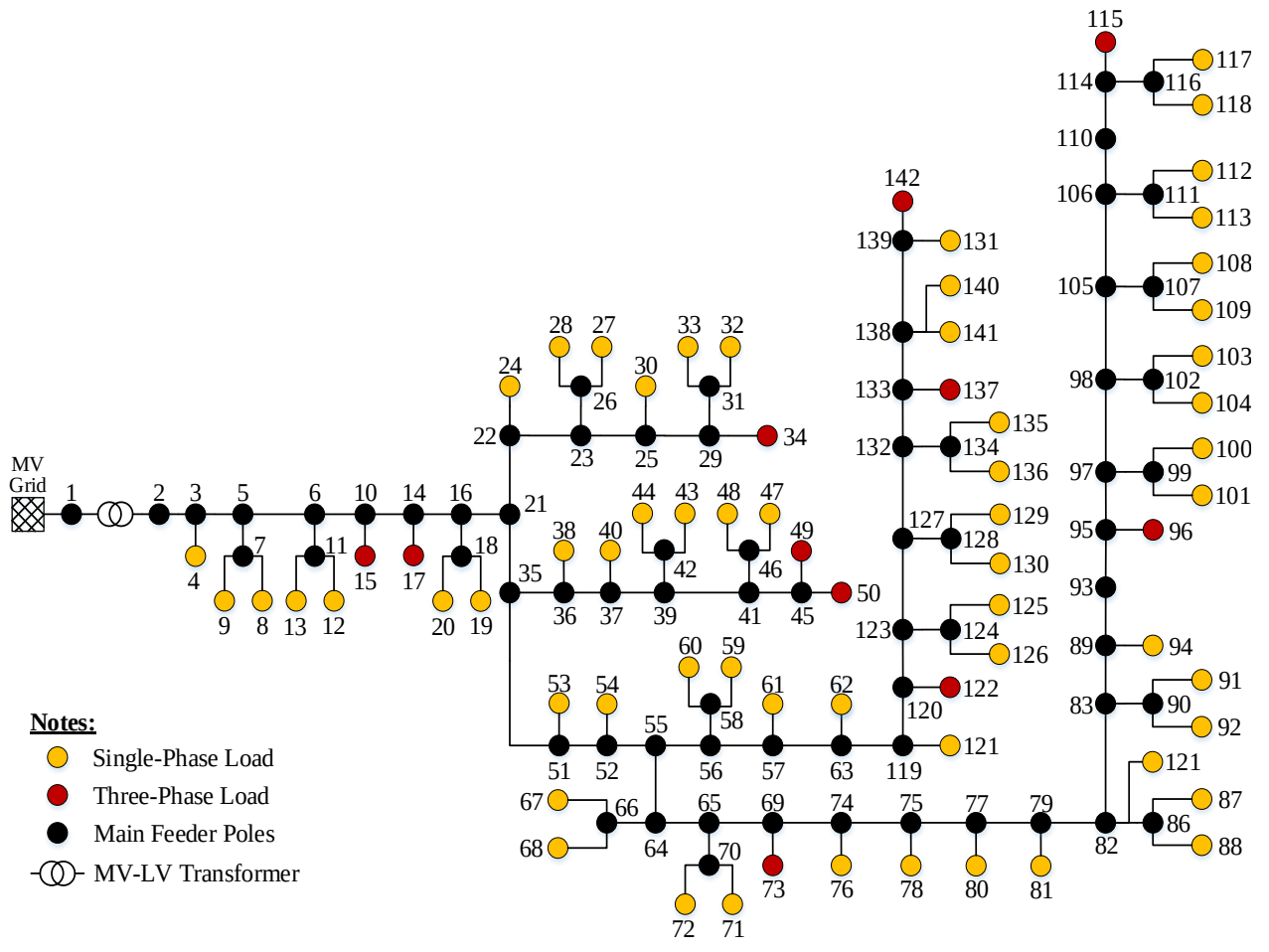

Fig. 2. IEEE European LVDG

The connection with the MVDG is achieved through an 800 kVA D-Y$_g$ 11 kV / 416 V transformer, located between nodes 1 and 2. Moreover, the neutral conductor is grounded only at the transformer within the substation. The accuracy of the C-WLS SE is evaluated under different loading conditions, as a percentage of the transformer's nominal power and at a 0.95 lagging power factor, as well as for different line lengths in order to investigate how the results are affected by different types of LVDGs (lightly loaded – highly loaded, urban - rural). Furthermore, due to the random nature of the measurement errors, the analysis is conducted in a Monte Carlo fashion in order to unbias the results, a small measurement error will yield more accurate results which can lead to inaccurate conclusions. Each scenario is executed 500 times and the average maximum voltage error, average voltage error and average phase angle error are calculated as,

$$\|\Delta V\|_{\infty}^{avg} = \frac{1}{MC} \sum_{i=1}^{MC} \max_{1 \le j \le N_{|V|}} \left\{ \left| \hat{V}_j^i - V_j^i \right| \right\} \tag{20}$$

$$\|\Delta V\|_{2}^{avg} = \frac{1}{MC} \sum_{i=1}^{MC} \frac{\sqrt{\sum_{j=1}^{N_{|V|}} \left| \hat{V}_j^i - V_j^i \right|^2}}{\sqrt{N_{|V|}}} \tag{21}$$

$$\|\Delta \theta\|_{2}^{avg} = \frac{1}{MC} \sum_{i=1}^{MC} \frac{\sqrt{\sum_{j=1}^{N_{\theta}} \left| \hat{\theta}_j^i - \theta_j^i \right|^2}}{\sqrt{N_{\theta}}} \tag{22}$$

where MC is the number of Monte Carlo iterations, $\hat{V}_j^i$ is the estimated value of the $j$-th voltage magnitude state variable at the $i$-th Monte Carlo iteration, $V_j^i$ is the true value of the corresponding state variable and $N_{|V|}$ is the total number of voltage magnitude state variables (similarly for $\hat{\theta}_j^i$, $\theta_j^i$ and $N_\theta$ for the average phase angle error).

In Figs. 3 and 4, the average maximum and average voltage errors as well as the average phase angle error are presented when the C-WLS SE is used for monitoring the test system, where $L$ denotes the base case with the original line lengths. Based on these figures, it can be concluded that the accuracy of the C-WLS SE, when applied for the monitoring of LVDGs with a neutral conductor grounded only at the substation, is highly dependent on the system loading conditions as well as the type of LVDG. The reason behind this dependency is the voltage drop across the neutral conductor, which increases with the loading conditions and the total line length of the system. As a result, rural systems (which have longer line length) are affected in a greater degree than urban systems as the loading of the system increases. This is due to the overall higher impedance of the neutral conductor in rural systems, which for the same amount of current flowing through it, it will create a higher voltage drop across the conductor compared to an urban system. As the voltage drop across the neutral conductor increases, the use of Kron's reduction (12) leads to a system model that is not representative of the actual physical system. Moreover, the C-WLS SE assumes that the voltage measurements are phase to ground measurements [5], [7].

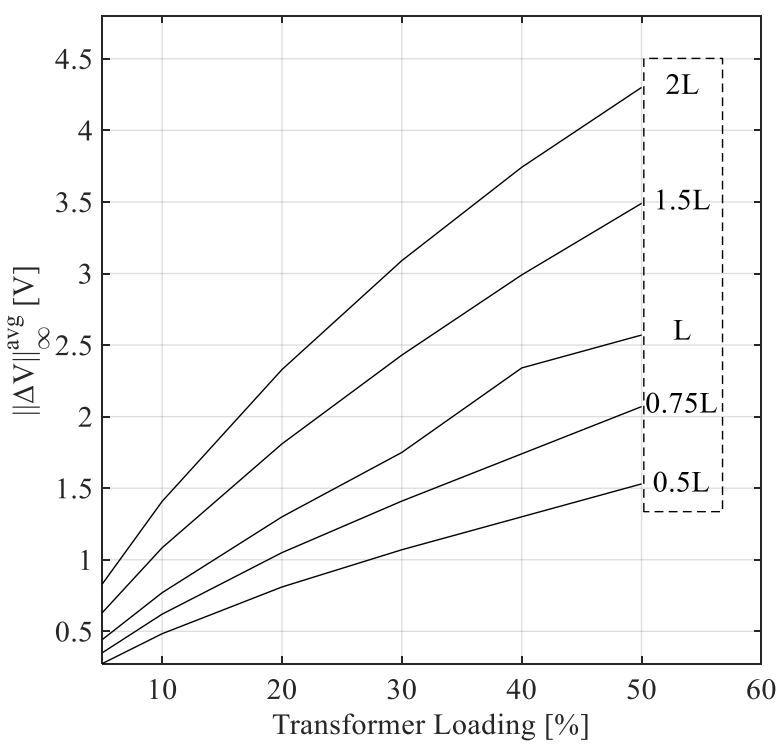

Fig. 3. Average maximum voltage error

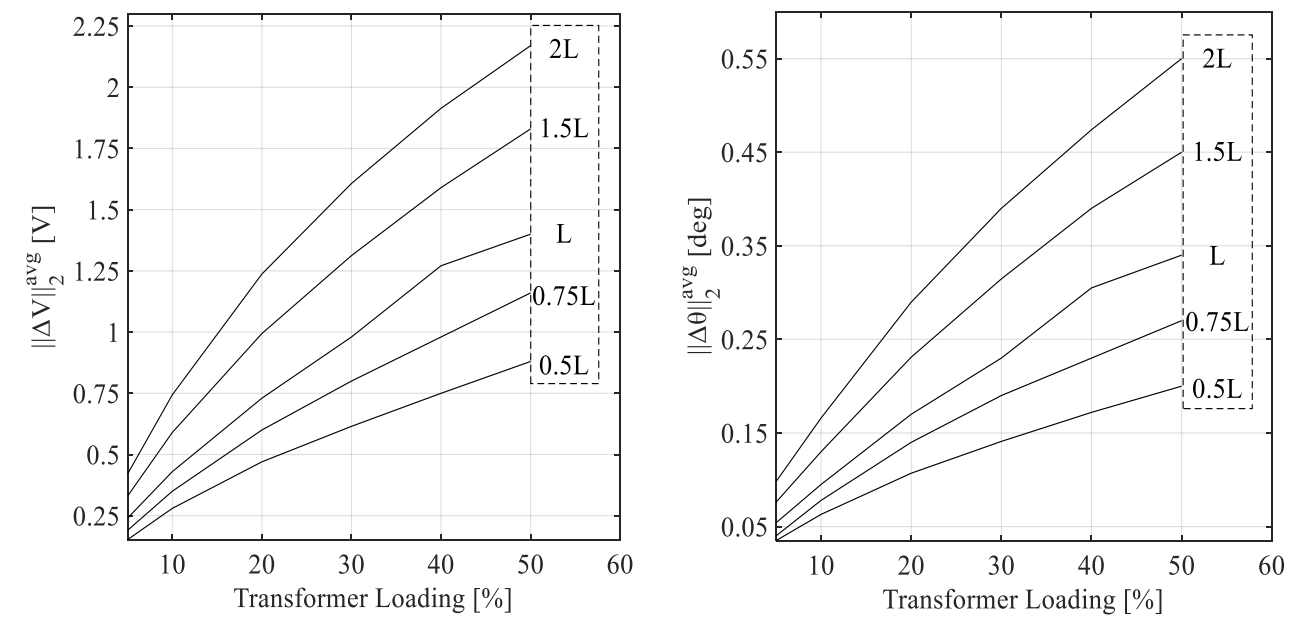

Fig. 4. Average voltage and phase angle errors

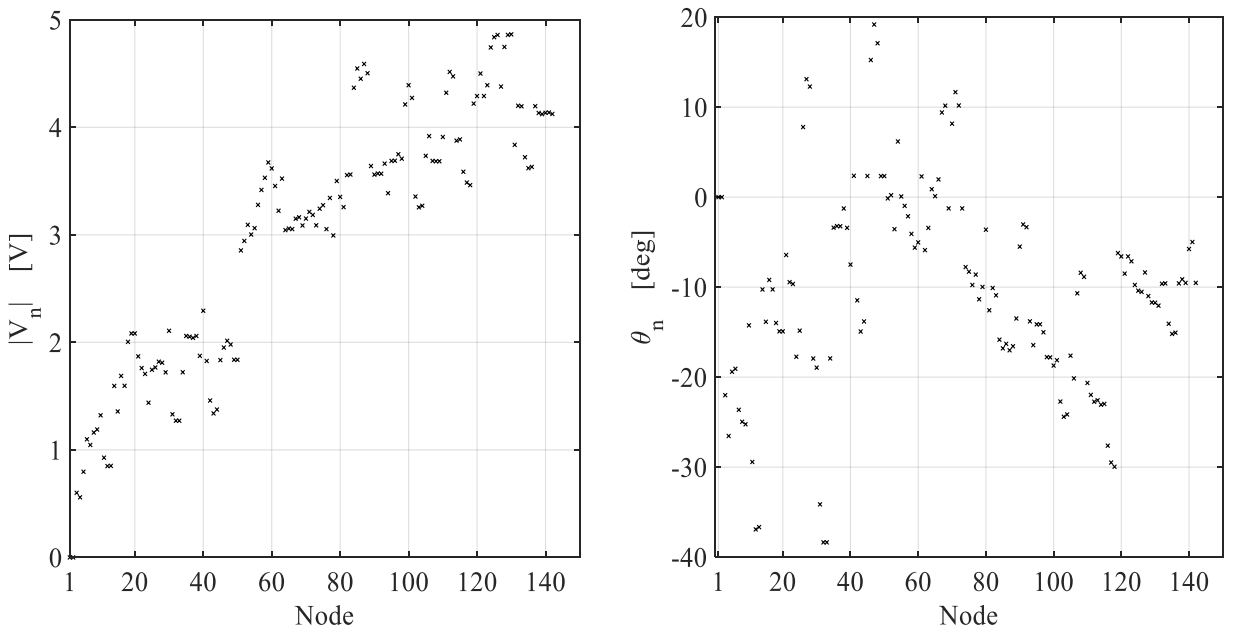

Fig. 5. Voltage of neutral conductor (50% loading and L line length)

However, smart meters provide phase to neutral measurements and since the voltage of the neutral conductor can have significantly different values across the system, as illustrated in Fig. 5, the measurements that are used in the C-WLS SE are not expressed under the same reference voltage. Therefore, the voltage measurements from the smart meters provide a distorted image of the system's condition which has a significant effect on the accuracy of the C-WLS SE.

## IV. WLS SE Considering The Neutral Conductor

In the previous section, the C-WLS SE methodology was presented and tested in order to evaluate its accuracy when used for the monitoring of a LVDG with a neutral conductor grounded only at the MV-LV substation. It was concluded that for this type of LVDGs, its accuracy was highly dependent on the loading conditions (mainly on the zero sequence current) for rural systems and in a lesser, but still significant, degree for urban systems. This proved that the C-WLS SE cannot be used as a general solution for the monitoring of any LVDG, but only under specific conditions. Moreover, the investigation results of the previous section have also highlighted the need for appropriate monitoring schemes that take into consideration the full characteristics of LVDGs. In this section, a new methodology, referred hereafter as N-WLS SE, is presented. The proposed methodology takes into consideration the full characteristics of LVDGs and as a result, the N-WLS SE exhibits significantly more accurate and consistent results compared to the C-WLS SE, without requiring any additional information about the system.

### A. N-WLS SE

In order to take into consideration the effect of the neutral conductor, the system is modeled as a four phase system where the neutral voltage is included in the state vector. Hence, every three phase node, excluding the reference node, in the system is characterized by eight state variables and every single phase node by four state variables, as given by,

$$x_{abcn} = (\theta_a, \theta_b, \theta_c, \theta_n, |V_a|, |V_b|, |V_c|, |V_n|) \quad (23)$$

$$x_{pn} = \left(\theta_p, \theta_n, |V_p|, |V_n|\right) \quad (24)$$

where $p = a, b, c$. This allows using the full 4x4 impedance matrix (9), therefore the use of Kron's reduction is not any longer necessary. As a result, the accuracy of the system's model is not affected by the state of the neutral voltage. Consequently, the measurement functions (5) and (6) are modified in order to include the neutral voltage in the calculations,

$$P_i^p = |V_i^p| \sum_{k=1}^{N} \sum_{m=a}^{n} |V_k^m| \left(G_{ik}^{pm} cos\delta_{ik}^{pm} + B_{ik}^{pm} sin\delta_{ik}^{pm}\right) \quad (25)$$

$$Q_i^p = |V_i^p| \sum_{k=1}^{N} \sum_{m=a}^{n} |V_k^m| \left(G_{ik}^{pm} sin\delta_{ik}^{pm} - B_{ik}^{pm} cos\delta_{ik}^{pm}\right) \quad (26)$$

where $m = a, b, c, n$ corresponds to the system phases and $N$ is the total number of nodes in the system. However, the new state variables that represent the neutral voltage in the system are unobservable since only measurements that can be related with the power phases are available [5]. As a result, the N-WLS SE may converge in a significantly different state from the true system state, as there is no available information to guide the estimation of the neutral's voltage. Furthermore, due to the coupling between the phases and the neutral, this will also affect the estimation of the phases' state. In order to overcome this obstacle, the monitoring scheme illustrated in Fig. 6 is proposed where its main steps are as follows:

1) *Virtual measurements for the voltage of the neutral conductor:* When there are updated measurements from the smart meters, first a power flow analysis is conducted using the active and reactive power injection measurements as well as the $Z_{matrix.}$ The power flow analysis is conducted with the use of the backwards/forwards sweep method [23], which is applicable for radial distribution networks. In the case of weakly meshed distribution networks, the method described in [27] can be used for the power flow analysis. Furthermore, since the active/reactive power injection measurements that are available were calculated by the smart meters using a phase to neutral voltage, the calculation of injected currents during the backwards sweep is adjusted accordingly,

$$I_i^p = \left(\frac{S_i^p}{V_i^p - V_i^n}\right)^* \quad (27)$$

where $I_i^p$ and $S_i^p$ are the injected current and apparent power respectively in node $i$ and phase $p$ and $V_i^p - V_i^n$ is the voltage phasor associated with the given apparent power measurement ($V_i^p$ is initialized at $1pu\measuredangle\{0°, -120°, 120°\} + phase\ shift$ and $V_i^n$ is initialized to zero, where $phase\ shift$ is the phase shift introduced by the transformer on the low voltage side in relation to the high voltage side).

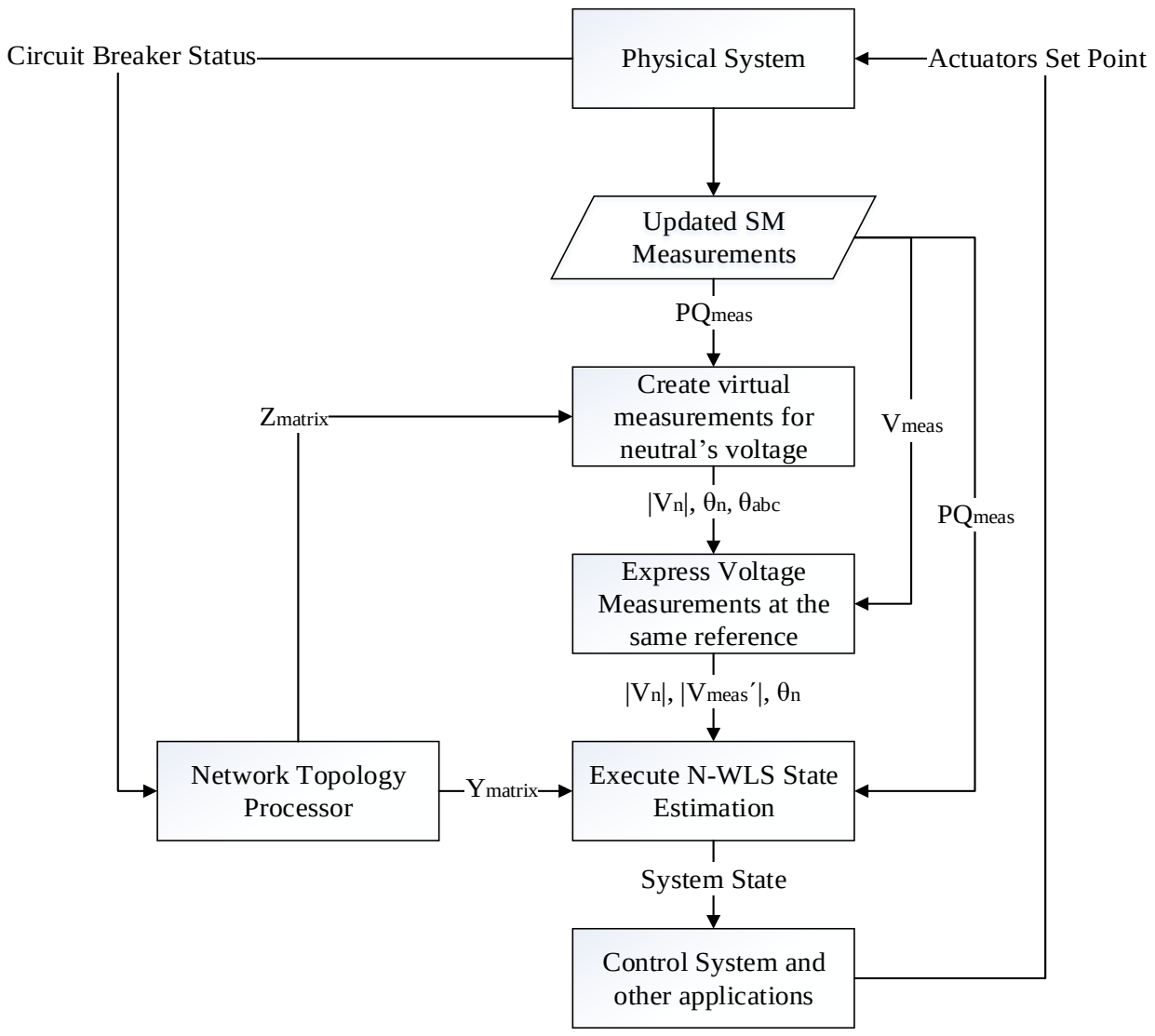

Fig. 6. Proposed monitoring scheme for LVDGs

2) *Inclusion of the neutral's virtual measurements in the measurement vector $h(x)$:* From the results of the power flow analysis, the neutral voltage magnitude and phase angle can be used as virtual measurements in the N-WLS SE, thus making the state variables that are related to the neutral voltage observable. The measurement function vector $h(x)$ and its Jacobian matrix $H(x)$ have the below format,

$$h(x) = \begin{pmatrix} P_{inj}(x) \\ Q_{inj}(x) \\ |V| \\ \theta_n \end{pmatrix} \quad (28)$$

$$H(x) = \begin{pmatrix} \frac{\partial P_{inj}}{\partial \theta_j} & \frac{\partial P_{inj}}{\partial V_k} \\ \frac{\partial Q_{inj}}{\partial \theta_j} & \frac{\partial Q_{inj}}{\partial V_k} \\ 0 & \frac{\partial V}{\partial V_k} \\ \frac{\partial \theta_n}{\partial \theta_j} & 0 \end{pmatrix} \quad (29)$$

where $|V|$ consists of the smart meters' voltage magnitude measurements as well as the neutral's voltage magnitudes calculated by the power flow analysis, $\theta_n$ is the neutral's phase angle across the system, $j = 1,..,N_\theta$ and $k = 1,..,N_{|V|}$. It should be noted that the measurement weights for the virtual measurements of the neutral voltage are implicitly chosen to be equal as the smart meters voltage measurements for the neutral's voltage magnitude and ten times higher for the virtual measurements of the neutral's phase angle due to their criticality. Furthermore, although the system state is available from the power flow analysis, the modified WLS SE must be applied in order to reduce the measurement noise and to improve the accuracy of the results. Since in the power flow analysis the measurement error is not accounted for, the results are heavily biased by the measurement devices. Moreover, by applying the modified WLS SE, bad data detection is enabled where erroneous measurements can be replaced by pseudo-measurements [5].

3) *Expressing all smart meters' voltage measurements under the same reference voltage:* Before the N-WLS SE is executed, first the calculated neutral voltage is used to express all voltage magnitude measurements from the smart meters under the same reference voltage,

$$|V'_{meas_i}| = \left| |V_{meas_i}| \measuredangle\theta_i + |V_{n_i}| \measuredangle\theta_{n_i} \right| \quad (30)$$

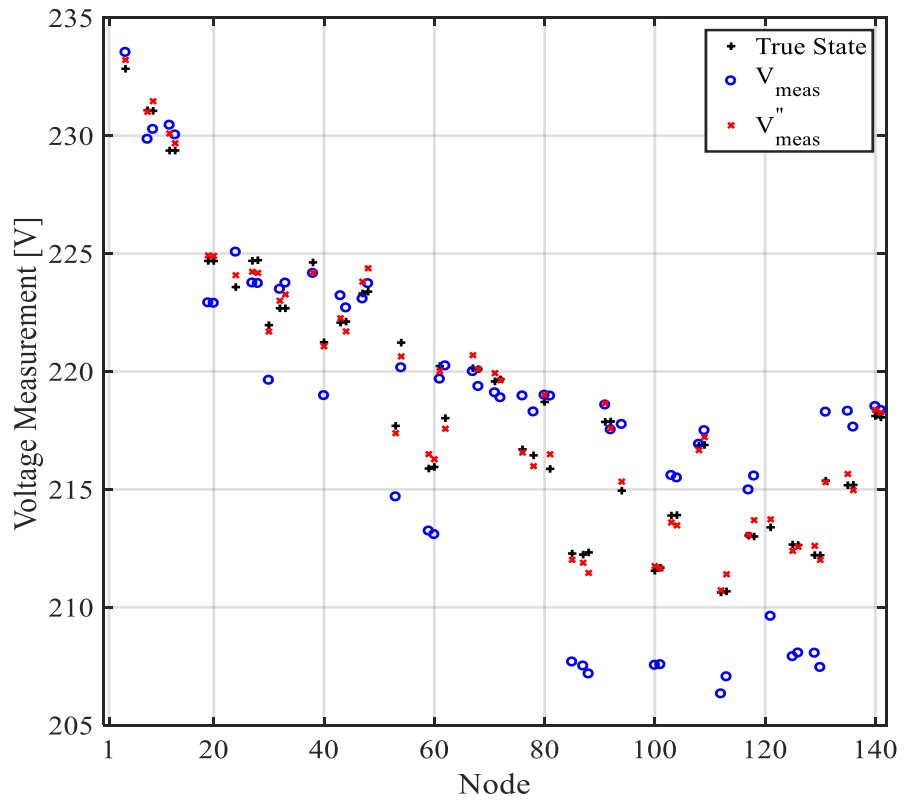

Fig. 7. Smart meter voltage measurements before and after their adjustment based on the calculated neutral voltage (50% loading and L line length)

where $|V_{meas_i}|$ is the $i$-th voltage magnitude measurement from the smart meters, $|V_{n_i}|$ and $\theta_{n_i}$ is the neutral voltage associated with the specific measurement, $\theta_i$ is the phase angle of the voltage measurement ($\theta_{n_i}$, $|V_{n_i}|$ and $\theta_i$ are provided by the power flow analysis) and $|V'_{meas_i}|$ is the corrected voltage magnitude measurement that is used in the N-WLS SE. The significance of this step is illustrated in Fig. 7 where the corrected voltage measurements are significantly closer to the true state of the corresponding system node compared to the unprocessed voltage measurements from the smart meters.

## B. Comparison with C-WLS

In Fig. 8, the mean of the absolute errors as well as their standard deviations for the average maximum voltage error $\|\Delta V\|_\infty$, average voltage $\|\Delta V\|_2$, and average phase angle $\|\Delta\theta\|_2$ errors are presented for both C-WLS and N-WLS. Based on this figure it can be concluded that the proposed monitoring scheme exhibits significantly better accuracy levels, especially for the estimation of the voltage magnitudes. Under the worst case for C-WLS, which is a rural high loaded system, N-WLS offers an improvement of 89% in the average maximum voltage error, 92.5% improvement in the average voltage error, and 81% improvement in the average phase angle error. As for as the phase angles are concerned, their estimation error for both methodologies and under all scenarios is insignificant in the overall operation of LVDGs and it can be concluded that the estimation of voltage magnitudes is the differentiating factor between C-WLS and N-WLS. Additionally, unlike C-WLS, N-

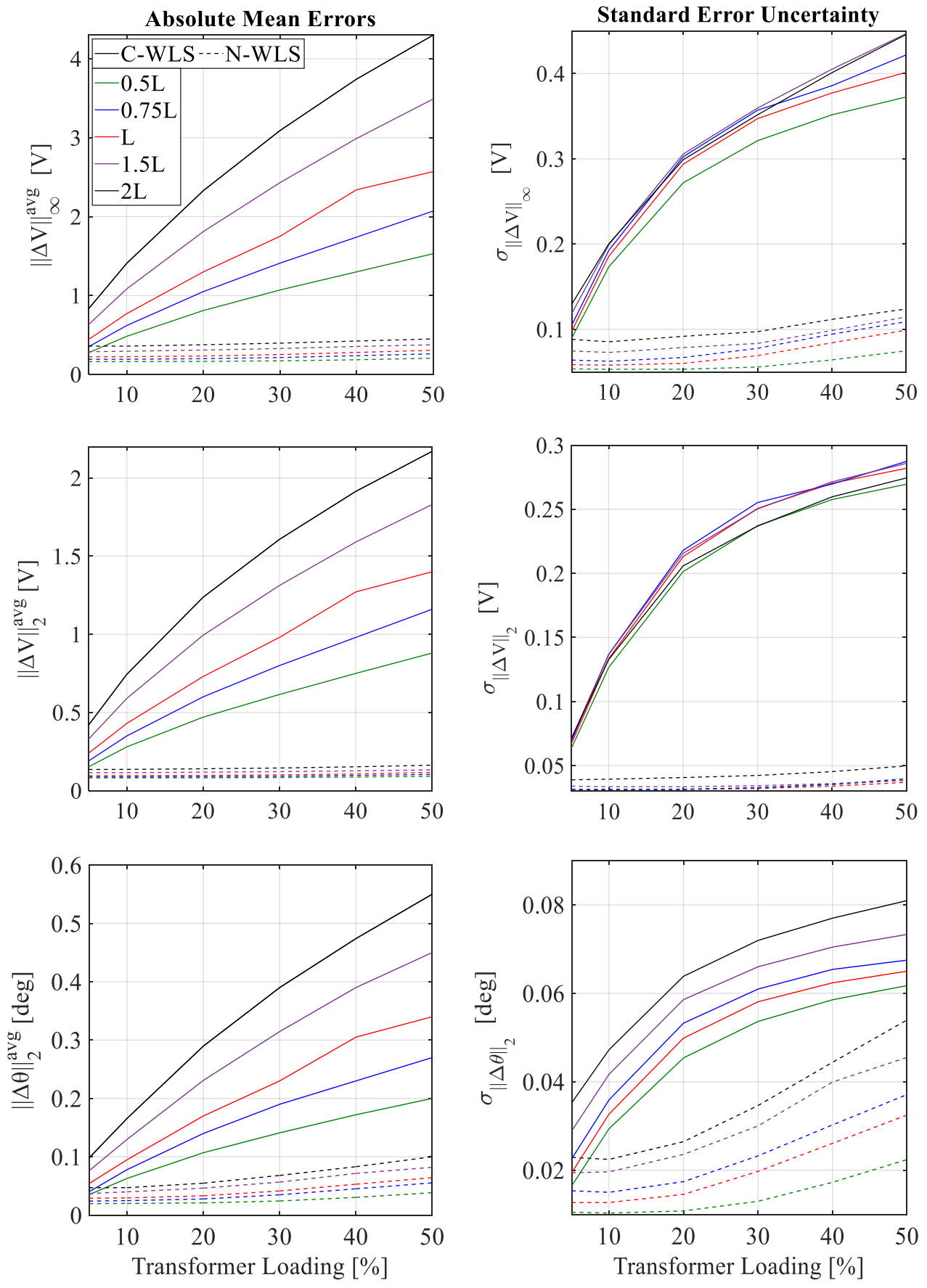

Fig. 8. Mean and standard deviation of absolute errors of C-WLS, N-WLS

WLS exhibits a relative constant performance, regarding the mean errors, across all scenarios, regardless of the loading conditions and type of LVDG. Moreover, from the standard deviation of the errors, it can be concluded that N-WLS is highly more consistent and hence more reliable compared to C-WLS. These results show that N-WLS SE can be used for the monitoring of any possible LVDG and under any loading conditions with unaffected performance, which is a highly important advantage over C-WLS SE. However, it must be mentioned that due to the increased complexity (increased number of state variables and the power flow analysis, which is an iterative process), the average computational time of N-WLS SE is almost twice as much as the computational time of C-WLS SE. Although this is a significant increase in the overall computational time, if it is considered that in reality, the process will have to run every 10-20 minutes due to the slow updating rate of smart meters, the proposed method can easily run in real-time applications. Hence, the improvement that the N-WLS SE offers in the accuracy of the estimated states outweighs the drawback of increased computational time and the proposed scheme can be used to achieve an accurate and reliable monitoring system for LVDGs.

## V. Conclusions

In this paper, the conventional methodology (C-WLS SE) for monitoring LVDGs was investigated regarding its performance when the neutral conductor of the system is grounded only at the MV-LV substation. The results of this investigation indicate that the performance of the C-WLS SE depends on both the loading conditions and type of LVDG, making its use only suitable for lightly loaded urban systems. Additionally, a new methodology was proposed that takes into consideration the full effect of the neutral conductor and as a result exhibits consistenly higher accuracy levels compared to the C-WLS SE without needing additional external information. Furthermore, as it can be concluded by the numerical simulations, the voltage magnitude and angle errors in the N-WLS SE method are relatively constant, regardless of the loading conditions and type of LVDG, illustrating the capabilities of this methodology as a general solution to the problem of LVDG monitoring. As LVDGs keep expanding and becoming more active with high penetration of DERs, their reliable and accurate monitoring will be of paramount importance. This paper has shown the deficiency of the C-WLS SE, which was developed for MVDGs, when applied in the monitoring of LVDGs, while simultaneously proposing a new methodology that is tailored to the characteristics of LVDGs.